\documentclass[aps,a4paper,superscriptaddress,nofootinbib,twocolumn,floatfix]{revtex4-1}
\usepackage{amsmath,amssymb,gensymb,graphicx,multirow,dcolumn,bm,latexsym,soul,nicefrac,acronym}
\graphicspath{{figures/}{../figures/}{../../figures/}}
\usepackage[mathscr]{euscript}
\usepackage{appendix}
\usepackage[colorlinks,linkcolor=blue,citecolor=blue,urlcolor=blue]{hyperref}
\usepackage{verbatim}
\usepackage{color}
\usepackage{subfigure}
\usepackage{longtable}
\usepackage{footnote}
\usepackage{cancel}
\usepackage{tikz}
\usetikzlibrary{arrows,positioning}
\usepackage[normalem]{ulem}
\newcommand{\HNU}{School of Physics, Henan Normal University, Xinxiang 453007, China}

\newcommand{\dd}{\mathrm{d}}

\newcommand{\BBH}{\mathrm{BBH}}

\begin{document}
\title{Forecasting Strongly Lensed Binary Black Hole Rates from the Stochastic Gravitational-Wave Background with TianQin}
\author{Zheng-Cheng Liang}
\email{liangzhengcheng@htu.edu.cn}
\affiliation{\HNU}
\date{\today}
\begin{abstract}
The millihertz stochastic gravitational-wave background (SGWB) provides a population-level probe of stellar-mass binary black holes (BBHs) and their cosmic merger history. We develop a framework using TianQin alone to propagate SGWB-informed BBH population constraints into forecasts of strongly lensed BBH mergers detectable by ground-based gravitational-wave detectors. The framework distinguishes between lensed events with at least one detected image and lens systems in which both images are detected and successfully associated. Using a one-year mock observation, we infer the BBH population and propagate the resulting posterior through a galaxy-scale lensing model with image-level selection. For the adopted system-magnification threshold, \(\mu_{\rm sys}\ge 2\), the median predicted rate of lensed events with at least one detected image is \(R_{\rm event}^{\rm det}=5.50\times10^{-4}\,\mathrm{yr}^{-1}\), while the median rate of associated two-image lens systems is \(R_{\rm sys}^{\rm det}=2.54\times10^{-4}\,\mathrm{yr}^{-1}\); the median detectable system fraction is \(\mathcal{F}_{\rm sys}^{\rm det}=3.35\times10^{-4}\). Compared with the prior-only propagation, the pooled 16th--84th percentile widths decrease by 6.5\% for the event rate, 6.4\% for the system rate, and 4.0\% for the system fraction. Our work establishes a quantitative link between millihertz SGWB inference and forecasts of strongly lensed BBH mergers for ground-based gravitational-wave observations.
\end{abstract}
\maketitle
\acrodef{SGWB}{stochastic gravitational-wave background}
\acrodef{GW}{gravitational wave}
\acrodef{CBC}{compact binary coalescence}
\acrodef{MBHB}{massive black hole binary}
\acrodef{SBBH}{stellar-mass binary black hole}
\acrodef{EMRI}{extreme-mass-ratio inspiral}
\acrodef{DWD}{double white dwarf}
\acrodef{BH}{black hole}
\acrodef{NS}{neutron star}
\acrodef{BNS}{binary neutron star}
\acrodef{LIGO}{Laser Interferometer Gravitational Wave Observatory}
\acrodef{TQ}{TianQin}
\acrodef{KAGRA}{Kamioka Gravitational Wave Detector}
\acrodef{ET}{Einstein Telescope}
\acrodef{DECIGO}{DECi-hertz Interferometer Gravitational-wave Observatory}
\acrodef{CE}{Cosmic Explorer}
\acrodef{NANOGrav}{North American Nanohertz Observatory for Gravitational Waves}
\acrodef{ORF}{overlap reduction function}
\acrodef{ASD}{amplitude spectral density}
\acrodef{PSD}{power spectral density}
\acrodef{SNR}{signal-to-noise ratio}
\acrodef{TDI}{time-delay interferometry}
\acrodef{GR}{general relativity}
\acrodef{PBH}{primordial black hole}
\acrodef{SSB}{solar-system barycenter}
\acrodef{PTA}{pulsar timing array}

\section{Introduction}\label{sec:intro}

The stochastic gravitational-wave background (SGWB) provides population-level information about compact-binary mergers that complements catalogs of individually resolved gravitational-wave (GW) events \cite{Maggiore2000,AllenRomano1999,Phinney2001}. Its spectrum integrates over source masses and redshifts, making it sensitive to the cosmic merger history while generally leaving degeneracies among the underlying population parameters \cite{Maggiore2000,AllenRomano1999,Phinney2001,Regimbau2011,Christensen2018,Caprini2019}. A useful forecast therefore propagates these population uncertainties into the observable of interest rather than treating a measured background amplitude as a unique reconstruction of the source history.

Planned space-based detectors such as TianQin, the Laser Interferometer Space Antenna (LISA), and Taiji target the millihertz band and provide complementary opportunities to study compact-binary populations and other GW sources \cite{Luo2016TianQin,Mei2021TQ,AmaroSeoane2017LISA,Ruan2020Taiji}. In this band, stellar-mass binary black holes (BBHs) contribute to the astrophysical SGWB~\cite{Zhu:2011bd,Inayoshi:2016hco,LIGOScientific:2016fpe,LIGOScientific:2017zlf,Yagi:2017zhb,Chen:2018rzo}, while unresolved Galactic double white dwarfs (DWDs) produce a foreground \cite{Bender:1997hs,Nelemans:2001hp,Barack:2004wc,Edlund:2005ye,Ruiter2010,Cornish:2017vip,Huang2020DWD,Liang2022TQSGWB,Liu:2023qap,Hofman:2024xar}. Extracting weak SGWB signal requires accurate modeling of both the detector response and instrumental noise. For space-based detectors, the constellation geometry evolves with time and the inter-spacecraft arm lengths are generally unequal, preventing laser-frequency noise from being canceled by a simple equal-arm Michelson combination. Time delay interferometry (TDI) addresses this problem by combining appropriately delayed inter-spacecraft phase measurements to suppress laser-frequency noise \cite{Tinto:1999yr,TintoDhurandhar2021}. From Michelson-like TDI observables, the commonly used noise-orthogonal \(A\), \(E\), and \(T\) channels can be constructed. The \(A\) and \(E\) channels retain substantial sensitivity to GWs, whereas the \(T\) channel is comparatively less sensitive to GWs at low frequencies \cite{Adams:2010vc,Caprini:2019pxz,Cheng:2022vct,Liang2023TDI,Baghi:2023qnq,Li:2024lvt}.

Strong lensing weights the same BBH population differently from the SGWB \cite{Dai2017Lensing,Oguri2018}. Magnification can raise a distant merger above a ground-based detection threshold, while the lensing optical depth and magnification distribution depend on source redshift and the lens population \cite{Dai2017Lensing,Oguri2018,LVC2021Lensing,LittenbergCornish2023}. A multiply imaged source may produce images with different magnifications, arrival times, and parities, and wave-optics effects can also matter in appropriate regimes \cite{Takahashi2003,Hannuksela2019,Li2018,Ng2018,Diego2019,Staelens2024}. Consequently, an observable lensing fraction depends on image-level selection and on whether the analysis counts detected images, source events, or associated lens systems. Here an event with at least one detected image is distinct from an associated lens system for which both images are detected and linked to the same source.

Previous studies have connected unresolved BBH backgrounds with the rates of individually detectable lensed mergers, while complementary work has developed population-based predictions for magnification bias and ground-based strong-lensing searches \cite{Oguri2018,Vitale2019,Buscicchio2020,Mukherjee2020Lensing,LVC2021Lensing,Lin:2026uyg}. Together, these studies establish the physical and observational basis for relating an unresolved binary population to a sample of resolved, magnified mergers. A natural next step is to propagate the population posterior inferred from a space-based SGWB measurement through the lensing and image-selection calculation, so that uncertainty in the background inference is reflected in the predicted rates for ground-based GW detectors.

We implement this programme for a single TianQin constellation in three linked steps. First, we infer the BBH population using the phase-dependent $2\times2$ complex spectral covariance of the $A/E$ TDI channels \cite{Liang2022TQSGWB,Liang2023TDI}, including their complex cross spectrum and the response-modulated Galactic DWD foreground \cite{Huang2020DWD,Banagiri2021} in a phase-folded complex Wishart likelihood \cite{Wishart1928,Goodman1963}. This ties the population posterior to a concrete detector-level measurement model rather than to a fixed SGWB amplitude \cite{Buscicchio2020,Mukherjee2020Lensing}. Second, we propagate that posterior through a galaxy-scale singular isothermal sphere (SIS) lens population and image-level detection and association rules \cite{Oguri2018,Dai2017Lensing,GerosaBellotti2024}, reporting separately the rate of lensed source events with at least one detected image and the rate of systems with both images detected and associated \cite{Buscicchio2020,Mukherjee2020Lensing}. Third, we compare posterior-predictive rates with prior-only propagation and test sensitivity to lens quadrature, frequency support, orbital-phase information, and the $A/E$ cross spectrum. Together, these steps connect a millihertz SGWB inference to explicitly defined lensing observables of ground-based detectors while keeping data-driven constraints distinct from assumptions imposed by the population and selection models.

The paper is organized as follows. Section~\ref{sec:model} defines the BBH population, the SIS lens population, and the image- and system-level observables. Section~\ref{sec:tqdata} describes the TianQin response, instrumental noise, and the Galactic DWD foreground. Section~\ref{sec:like} presents the phase-folded spectral likelihood and forecast configuration, and briefly summarizes mock-data generation and implementation checks. Section~\ref{sec:results} reports the rate forecasts and sensitivity tests. Section~\ref{sec:summary} summarizes the interpretation, limitations, and conclusions.

\section{Source population and strong-lensing selection model}\label{sec:model}

\subsection{Binary black hole population and stochastic background}

We characterize an isotropic stochastic background by the fractional energy density per logarithmic frequency interval, following the standard SGWB convention \cite{AllenRomano1999,Maggiore2000}:
\begin{equation}
	\Omega_{\rm GW}(f)\equiv\frac{1}{\rho_c}\frac{{\rm d}\rho_{\rm GW}}{{\rm d}\ln f},
	\qquad
	\rho_c=\frac{3H_0^2}{8\pi G}\quad(c=1),
	\label{eq:omega_def}
\end{equation}
where \(H_0\) is the Hubble constant. Throughout this subsection, the cosmological equations use geometric units, \(c=1\), so that \(\rho_c\) denotes the corresponding critical density. If an SI energy density is used instead, one restores \(\rho_{c,\mathrm{SI}}=3H_0^2c^2/(8\pi G)\), together with the appropriate factors of \(c\) in the source-energy relation. For conversion to the one-sided power spectral density (PSD), \(H_0\) is evaluated in SI units.

For a population of astrophysical sources, the background can be written in terms of the source-frame energy spectrum as \cite{Phinney2001,Regimbau2011,LIGOScientific:2016fpe,LIGOScientific:2017zlf}
\begin{equation}
	\begin{aligned}
		\Omega_{\rm GW}(f)=\frac{f}{\rho_c}
		\int {\rm d}\boldsymbol{\theta}\int_{0}^{z_{\rm max}} {\rm d}z\,
		\frac{R(z,\boldsymbol{\theta})}{(1+z)H(z)}
		\frac{{\rm d}E_{\rm GW}}{{\rm d}f_{\rm s}}(f_{\rm s},\boldsymbol{\theta}),
	\end{aligned}
	\label{eq:omega_population}
\end{equation}
where $R(z,\boldsymbol\theta)$ is the source-frame merger-rate density per comoving volume and source-frame time for population parameters $\boldsymbol\theta$, $\mathrm dE_{\rm GW}/\mathrm df_s$ is the source-frame GW energy spectrum, $f_s=(1+z)f$ is the source-frame frequency, and \(H(z)\) is the Hubble parameter. We integrate to $z_{\max}=10$ and use the flat-$\Lambda$CDM approximation $H(z)\simeq H_0\sqrt{\Omega_{\rm m}(1+z)^3+\Omega_\Lambda}$, where $\Omega_{\rm m}$ and $\Omega_\Lambda$ are the present-day matter and cosmological-constant density parameters.

In the millihertz regime, we adopt the inspiral spectrum for the stellar-mass BBH contribution \cite{Phinney2001,Regimbau2011}:
\begin{equation}
	\Omega_{\BBH}(f)=A_{\BBH}\left(\frac{f}{f_0}\right)^{\alpha_{\BBH}},
	\qquad f_0=3\,\mathrm{mHz},
	\label{eq:bbh_powerlaw}
\end{equation}
where $A_{\BBH}$ is the dimensionless amplitude at the reference frequency \(f_0\), and the index $\alpha_{\BBH}$ is a spectral-shape coordinate whose prior is centered on the circular-inspiral value \(2/3\) rather than fixed to it.

The redshift evolution used to map the measured amplitude to high-redshift merger rates is a phenomenological parameterization motivated by population analyses~\cite{Abbott2016GWTC,Abbott2021GWTC2,Abbott2021Population,Fishbach2018,Vitale2019,Sathyaprakash2009}:
\begin{equation}
	\begin{aligned}
		R_{\BBH}(z;\lambda_z,\alpha_z,z_p)
		&=R_{0}\,\frac{(1+z)^{\lambda_z}}
		{1+\left[(1+z)/(1+z_p)\right]^{\lambda_z+\alpha_z}}\\
		&\quad\times\left[1+\left(\frac{1}{1+z_p}\right)^{\lambda_z+\alpha_z}\right],
	\end{aligned}
	\label{eq:rate_model}
\end{equation}
where $R_0=R_{\BBH}(0)$ is the local source-frame rate density, $\lambda_z$ controls the low-to-intermediate-redshift growth, $\alpha_z$ controls the high-redshift decline, and $z_p$ sets the turnover redshift. In this work, we adopt Eq.~\eqref{eq:rate_model} with the high-redshift turnover active, fixing $\alpha_z=5.6$, $z_p=1.9$, and $\lambda_z=2.5$. The normalization factor in Eq.~\eqref{eq:rate_model} enforces $R_{\BBH}(0)=R_0$, and the turnover avoids the unbounded high-redshift growth of a pure power law. The PixelPop event-rate distribution of Gravitational-Wave Transient Catalog 5.0 (GWTC-5.0) is reported at \(z=0.2\)~\cite{LIGO2026GWTC5}, we rescale it to \(z=0\) using the normalized turnover kernel in Eq.~\eqref{eq:rate_model} with the same fixed \(\alpha_z\), \(z_p\), and \(\lambda_z\). This gives the local merger rate \(R_0=23.76\,\mathrm{Gpc}^{-3}\,\mathrm{yr}^{-1}\).

The corresponding mass-resolved rate is defined using the normalized effective chirp-mass distribution in the adopted population model \cite{LIGO2026GWTC5}:
\begin{equation}
	\begin{aligned}
		R_{\BBH}(z,\mathcal M|\boldsymbol\theta)&=R_{\BBH}(z|\boldsymbol\theta)\,p_{\mathcal M}(\mathcal M|z,\boldsymbol\theta),\\
		&\int\dd\mathcal M\,p_{\mathcal M}(\mathcal M|z,\boldsymbol\theta)=1,
	\end{aligned}
	\label{eq:mass_resolved_rate}
\end{equation}
where \(p_{\mathcal M}\) is the effective mass distribution from the PixelPop model of GWTC-5.0~\cite{LIGO2026GWTC5}.

The amplitude \(A_{\BBH}\) is fixed by the same rate and mass model. For circular Newtonian inspiral, with the mass density normalized under the adopted mass measure,
\begin{equation}
	A_{\BBH}(\boldsymbol\theta)=
	\frac{(\pi G)^{2/3}f_0^{2/3}}{3\rho_c}
	\int_0^{z_{\max}}\!\dd z\,
	\frac{R_{\BBH}(z|\boldsymbol\theta)\,
		\langle\mathcal M^{5/3}\rangle_z}
	{(1+z)^{4/3}H(z)},
	\label{eq:bbh_amplitude_population}
\end{equation}
where \(\langle\mathcal M^{5/3}\rangle_z=\int\dd\mathcal M\,p_{\mathcal M}(\mathcal M|z,\boldsymbol\theta)\mathcal M^{5/3}\). Thus the direct spectral amplitude and the population grid are two representations of one physical quantity rather than independent parameters. \(A_{\BBH}\) is induced by the population model, and $\alpha_{\BBH}$ controls the measured spectral tilt. The explicit amplitude parameterization used in auxiliary cross-band diagnostics is a reparameterization of the same spectrum, with its population normalization kept consistent with Eq.~\eqref{eq:bbh_amplitude_population}.

The reported observer-frame all-sky rate above a threshold \(z_*\) uses the standard conversion from source-frame to observer-frame time \cite{Abbott2019BBHRates,Abbott2021Population,Fishbach2018}:
\begin{equation}
	\mathcal R_{\BBH}(z>z_*)=
	\int_{z_*}^{z_{\max}}
	\frac{R_{\BBH}(z)}{1+z}
	\frac{\dd V_c}{\dd z}\,\dd z ,
	\label{eq:highz_rate}
\end{equation}
where the factor \((1+z)^{-1}\) converts source-frame time to observer time. This is an observer-frame all-sky merger rate. We use \(z_*=2\) and \(z_*=4\) as diagnostic summaries of the high-redshift population; they are not the primary observable of this paper, which is the selection-weighted fraction introduced in the main text.

Because the power-law spectrum supplies only a single dominant amplitude integral, the SGWB alone generally cannot separate the redshift-evolution and mass-model directions that enter \(A_{\BBH}(\boldsymbol\theta)\). The physical lensing calculation below therefore propagates the full population posterior rather than identifying a unique redshift history from the background amplitude.
\subsection{Singular isothermal sphere lens population and image properties}

We model the strong-lensing population with galaxy-scale SISs, following the standard galaxy-lens rate treatment of Ref.~\cite{Oguri2018}. The velocity-dispersion function is
\begin{equation}
	\frac{\dd n_{\rm lens}}{\dd\sigma_v}
	=\phi_*\frac{\beta}{\sigma_*\,\Gamma(\alpha/\beta)}
	\left(\frac{\sigma_v}{\sigma_*}\right)^{\alpha-1}
	\exp\left[-\left(\frac{\sigma_v}{\sigma_*}\right)^\beta\right],
	\label{eq:sis_vdf}
\end{equation}
with fixed parameters $\phi_*=8.0\times10^{-3}\,\mathrm{Mpc}^{-3}$, $\sigma_*=161\,\mathrm{km\,s^{-1}}$, $\alpha=2.32$, and $\beta=2.67$, over the interval $80\leq\sigma_v/(\mathrm{km\,s^{-1}})\leq400$. This is a conditional lens-population model.

Over the finite interval used here, the velocity-dispersion function is normalized as
\begin{equation}
	\begin{aligned}
		\int_{\sigma_{\min}}^{\sigma_{\max}}\frac{\dd n_{\rm lens}}{\dd\sigma_v}\,\dd\sigma_v
		&=\phi_*\frac{\Delta\gamma}{\Gamma(\alpha/\beta)},\\
		\Delta\gamma&\equiv\gamma\!\left[\alpha/\beta,x_{\max}^{\beta}\right]-\gamma\!\left[\alpha/\beta,x_{\min}^{\beta}\right],\\
		x_{\min,\max}&=\frac{\sigma_{\min,\max}}{\sigma_*},
	\end{aligned}
	\label{eq:sis_vdf_integral}
\end{equation}
where $\gamma$ is the lower incomplete gamma function and $\Gamma$ is the Euler gamma function. Thus $\phi_*$ is the parent comoving density, while the finite integration limits determine the density entering the optical depth.

For a source at $z_s$ and a lens at $z_l$, the standard SIS lens equation gives the multiply imaged region $0<y<1$ \cite{Oguri2018,Dai2017Lensing}. Here $y$ is the source-plane impact parameter in units of the Einstein angle, with probability measure $p(y)\,\dd y=2y\,\dd y$. The absolute magnifications of the two images are
\begin{equation}
	|\mu_+|=1+\frac{1}{y},\qquad |\mu_-|=\frac{1}{y}-1,
	\label{eq:sis_magnifications}
\end{equation}
and the negative parity of the second image is retained in the image bookkeeping. The corresponding SIS arrival-time difference is \cite{Oguri2018,Dai2017Lensing}
\begin{equation}
	\Delta t_{+-}=32\pi^2\left(\frac{\sigma_v}{c}\right)^4
	\frac{D_lD_{ls}}{D_s}(1+z_l)\frac{y}{c},
	\label{eq:sis_delay}
\end{equation}
where $c$ is the speed of light, and $D_l$, $D_s$, and $D_{ls}$ are the angular-diameter distances from observer to lens, observer to source, and lens to source, respectively. In the two-image SIS region, the system magnification is $\mu_{\rm sys}=|\mu_+|+|\mu_-|=2/y$. It is used only for thresholding; image selection continues to depend on each image separately.

The lensing probability measure follows from the SIS optical-depth element
\begin{equation}
	\begin{aligned}
		\dd P_{\rm lens}&=(1+z_l)^2\frac{c}{H(z_l)}
		\pi[D_l\theta_E(z_l,z_s,\sigma_v)]^2\\
		&\quad\times \frac{\dd n_{\rm lens}}{\dd\sigma_v}\,\dd z_l\,\dd\sigma_v,\\
		\theta_E&=4\pi\left(\frac{\sigma_v}{c}\right)^2\frac{D_{ls}}{D_s},
	\end{aligned}
	\label{eq:sis_optical_depth}
\end{equation}
with the source-plane impact-parameter measure $2y\,\dd y$ over $0<y<1$. Here $\theta_E$ is the Einstein angle; the remaining SIS parameters are defined as above.

\subsection{Image selection and event- and system-level rates}

For an unlensed source with source-frame chirp mass \(\mathcal M\) at redshift $z$, the detector-frame chirp mass is \(\mathcal M_z=(1+z)\mathcal M\). We adopt the following optimal signal-to-noise ratio (SNR) scaling as a fixed selection calibration; it is a simplifying assumption of this forecast rather than a pipeline-calibrated detection efficiency \cite{GerosaBellotti2024}:
\begin{equation}
	\begin{aligned}
		\rho_0(z,\mathcal M)&=8\left(\frac{\mathcal M_z}{26M_\odot}\right)^{5/6}
		\frac{D_L(0.1)}{D_L(z)},\\
		\rho_i(\omega)&=\omega\sqrt{|\mu_i|}\,\rho_0,
	\end{aligned}
	\label{eq:image_snr}
\end{equation}
where $\rho_0$ is the unlensed optimal SNR under the adopted calibration, $\omega$ is the orientation-dependent projection factor, $D_L(z)$ is the luminosity distance, and $D_L(0.1)$ is the reference distance at redshift $z=0.1$. For a fixed sky position, polarization, and inclination, we evaluate image detectability with the noncentral-chi-square survival function, equivalently the first-order Marcum-$Q$ function:
\begin{equation}
	P_i=Q_1(\rho_i,\rho_{\rm th}),\qquad \rho_{\rm th}=8,
	\label{eq:marcum_q}
\end{equation}
where \(Q_1\) is the first-order Marcum-\(Q\) function, the survival function associated with the noncentral-\(\chi^2\) detection statistic.

We average over isotropic sky position $(\theta,\varphi)$, polarization $\psi$, and inclination $\iota$ using the standard isotropic orientation measure adopted in lens-selection forecasts \cite{Oguri2018,GerosaBellotti2024}:
\begin{equation}
	\begin{aligned}
		\langle X\rangle_\omega
		=&\frac{1}{8\pi^2}\int_{-1}^{1}\dd(\cos\theta)
		\int_{0}^{2\pi}\dd\varphi
		\int_{0}^{\pi}\dd\psi\\
		&\times\int_{-1}^{1}\dd(\cos\iota)\,X(\theta,\varphi,\psi,\iota).
	\end{aligned}
	\label{eq:orientation_average}
\end{equation}
We evaluate this integral with a scrambled Sobol sequence. The same orientation point is used for both images, so the shared binary orientation is preserved.

Since the two images share the same binary orientation, we combine their image-level detection probabilities before averaging over the common orientation variable \(\omega\), following the usual source-level lens-selection bookkeeping \cite{Oguri2018,Buscicchio2020}:
\begin{equation}
	\begin{aligned}
		P_{\rm event}&=\left\langle1-(1-P_+)(1-P_-)\right\rangle_{\omega},\\
		P_{\rm sys}&=\left\langle P_+P_-\right\rangle_{\omega}\,
		\epsilon_{\rm assoc}\,\mathbf 1(\Delta t_{+-}<T_{\rm assoc}),
	\end{aligned}
	\label{eq:event_system_selection}
\end{equation}
where $\mathbf 1(\mathcal A)$ denotes the indicator function, which equals one if condition $\mathcal A$ is true and zero otherwise; thus $\mathbf 1(\Delta t_{+-}<T_{\rm assoc})$ enforces the image-association time window. In the baseline calculation we take \(\epsilon_{\rm assoc}=1\) and \(T_{\rm assoc}=365.25\,\mathrm d\). Here \(T_{\rm assoc}\) is the time window within which two images are treated as one associated system. The one-year simulated data set has duration \(T_{\rm data}=365.25\,\mathrm d\). Although \(T_{\rm assoc}=T_{\rm data}\) in the baseline, \(T_{\rm assoc}\) enters only the image-association selection and is not a second Fourier-data duration.

Combining the source-frame-to-observer-frame rate conversion with the SIS lensing probability, we define the at-least-one-image event rate and the two-image system rate as \cite{Oguri2018,Buscicchio2020,Mukherjee2020Lensing}
\begin{equation}
	\begin{aligned}
		R_{\rm event}^{\rm det}(\mu_{\rm sys}\ge\mu_0)&=\int\dd z_s\,\dd\mathcal M\,
		\frac{R_{\BBH}(z_s,\mathcal M|\boldsymbol\theta)}{1+z_s}\\
		&\quad\times\frac{\dd V_c}{\dd z_s}\int\dd P_{\rm lens}\,P_{\rm event}
		\mathbf 1(\mu_{\rm sys}\ge\mu_0),\\
		R_{\rm sys}^{\rm det}(\mu_{\rm sys}\ge\mu_0)&=\int\dd z_s\,\dd\mathcal M\,
		\frac{R_{\BBH}(z_s,\mathcal M|\boldsymbol\theta)}{1+z_s}\\
		&\quad\times\frac{\dd V_c}{\dd z_s}\int\dd P_{\rm lens}\,P_{\rm sys}
		\mathbf 1(\mu_{\rm sys}\ge\mu_0).
	\end{aligned}
	\label{eq:physical_rates}
\end{equation}
Here $z_s$ is the source redshift, $\mathrm dV_c/\mathrm dz_s$ is the differential comoving volume, $\boldsymbol\theta$ denotes the BBH population parameters, $\mathrm dP_{\rm lens}$ is the differential SIS lensing probability defined in Eq.~\eqref{eq:sis_optical_depth}, and $\mu_0$ is the minimum total magnification used to select a system. The event rate requires at least one detected image; the system rate requires both images to be detected and associated.

The all-detected system rate counts each unlensed source and each successfully associated lensed system once,
\begin{equation}
	\begin{aligned}
		R_{\rm unlensed}^{\rm det}&=\int\dd z_s\,\dd\mathcal M\,
		\frac{R_{\BBH}(z_s,\mathcal M|\boldsymbol\theta)}{1+z_s}
		\frac{\dd V_c}{\dd z_s}\\
		&\quad\times[1-\tau_{\rm SIS}(z_s)]\,P_0(z_s,\mathcal M),\\
		R_{\rm lensed,sys}^{\rm det}&=\int\dd z_s\,\dd\mathcal M\,
		\frac{R_{\BBH}(z_s,\mathcal M|\boldsymbol\theta)}{1+z_s}
		\frac{\dd V_c}{\dd z_s}\\
		&\quad\times\int\dd P_{\rm lens}\,P_{\rm sys},\\
		R_{\rm all}^{\rm det}&=R_{\rm unlensed}^{\rm det}+R_{\rm lensed,sys}^{\rm det},\\
		\mathcal F_{\rm sys}^{\rm det}(\mu_0)&=
		\frac{R_{\rm sys}^{\rm det}(\mu_{\rm sys}\ge\mu_0)}{R_{\rm all}^{\rm det}}.
	\end{aligned}
	\label{eq:system_fraction}
\end{equation}
Here \(P_0=\langle Q_1(\rho_0\omega,\rho_{\rm th})\rangle_\omega\) is the unlensed image-selection probability, and \(\tau_{\rm SIS}(z_s)=\int\dd P_{\rm lens}\) is the SIS multiple-imaging optical depth.

For comparison with event-based catalogues, we also define
\begin{equation}
	\begin{aligned}
		\mathcal F_{\rm event}^{\rm det}(\mu_0)
		&=\frac{R_{\rm event}^{\rm det}(\mu_{\rm sys}\ge\mu_0)}
		{R_{\rm unlensed}^{\rm det}+R_{\rm lensed,event}^{\rm det}},\\
		\mathcal F_{\rm sys|event}^{\rm det}(\mu_0)
		&=\frac{R_{\rm sys}^{\rm det}(\mu_{\rm sys}\ge\mu_0)}
		{R_{\rm unlensed}^{\rm det}+R_{\rm lensed,event}^{\rm det}},
	\end{aligned}
	\label{eq:event_fraction}
\end{equation}
where \(R_{\rm lensed,event}^{\rm det}\) is the all-threshold lensed rate for which at least one image is detected. We use \(R_{\rm event}^{\rm det}\) for an event rate with at least one detected image, and \(R_{\rm sys}^{\rm det}\) for a system rate in which both images are detected and satisfy the association condition. The quantity \(\mathcal F_{\rm sys}^{\rm det}\) is a conditional system fraction with the system-counting denominator in Eq.~\eqref{eq:system_fraction}; it is neither a single-source lensing probability nor the fraction of all detected events that are lensed. The auxiliary \(\mathcal F_{\rm sys|event}^{\rm det}\) mixes system and event denominators and is therefore not the primary system fraction. A lensed source with only one detected image, or with two images that fail the association condition, contributes to the event rate when at least one image is detected but not to the system denominator. Consequently, each unlensed source and each associated lensed source is counted once, and image counts are never added to \(R_{\rm all}^{\rm det}\).

\section{Spectral covariance model}\label{sec:tqdata}

\subsection{TianQin configuration and TDI noise}

We model TianQin as an approximately equilateral geocentric constellation of three spacecraft with arm length $L_{\rm TQ}=\sqrt{3}\times10^8\,\mathrm m$. Its constellation plane and line of sight follow the nominal pointing configuration described in Refs.~\cite{Luo2016TianQin,Mei2021TQ}, and the detector response is evaluated along the corresponding analytic orbital motion. The constellation rotation period is $P_{\rm TQ}\simeq3.65$ days. These mission-geometry choices determine the time-dependent antenna response and, in particular, the periodic modulation of an anisotropic Galactic foreground.

The basic interferometric observables are formed from one-way inter-spacecraft phase measurements. TDI combines delayed links to cancel the dominant laser-frequency noise in the presence of unequal and slowly varying arms \cite{Prince2002,Tinto2005,Tinto2014}. In this work, we focus on the orthogonal $A/E$ TDI basis: the $A$ and $E$ channels carry the leading low-frequency GW response in the adopted frequency range, whereas the $T$ channel has a suppressed long-wavelength response. 

The diagonal TianQin instrumental PSDs are evaluated with the $A/E$ transfer function and convention described for TianQin in Ref.~\cite{Liang2023TDI},
\begin{equation}
	\begin{aligned}
		P_A(f)=P_E(f)&=\frac{2\sin^2u}{L_{\rm TQ}^2}\Bigg[(2+\cos u)S_p(f)\\
		&\quad+2\left(3+2\cos u+\cos 2u\right)\frac{S_a(f)}{(2\pi f)^4}\Bigg],\\
		u&=\frac{2\pi fL_{\rm TQ}}{c},\qquad L_{\rm TQ}=\sqrt{3}\times10^8\,\mathrm m.
	\end{aligned}
	\label{eq:prd107_noise}
\end{equation}
The reference PSD levels are $S_{p,0}=10^{-24}\,\mathrm{m^2\,Hz^{-1}}$ and $S_{a,0}=10^{-30}\,\mathrm{m^2\,s^{-4}\,Hz^{-1}}$, with $S_p(f)=S_{p,0}$ and $S_a(f)=S_{a,0}[1+10^{-4}\,\mathrm{Hz}/f]$. We set $N_{AA}=N_{EE}=P_A(f)$ and the cross-spectral density $N_{AE}=0$.

\subsection{Galactic foreground and covariance model}

The unresolved DWD population is the dominant astrophysical foreground included in this analysis. Population-synthesis studies predict a large number of slowly evolving binaries, producing a confusion foreground at low frequencies and a transition toward individually resolvable sources at higher frequencies \cite{Nelemans2001,Ruiter2010,Huang2020DWD,Korol2017}. We use the TianQin-specific one-year polynomial fit of Ref.~\cite{Huang2020DWD} to calibrate the Galactic foreground spectrum,
\begin{equation}
	\begin{aligned}
		\log_{10} h_{\rm DWD}^{(1\,\mathrm{yr})}(f)&=\sum_{i=0}^{6}a_i x^i,\\
		(a_0,\ldots,a_6)&=(-18.6,-1.13,-0.945,-1.02,4.05,\\
		&\quad -4.5,-0.5),
	\end{aligned}
	\label{eq:tianqin_dwd_fit}
\end{equation}
where $x\equiv\log_{10}[f/(1\,\mathrm{mHz})]$, corresponding to the one-year row of the fit table, and $h_{\rm DWD}^{(1\,\mathrm{yr})}$ is the fitted amplitude spectral density (ASD). Under the one-sided strain-PSD convention, we define the conversion factor $\omega_{\rm pref}(f)\equiv 2\pi^2 f^3/(3H_0^2)$. The reference energy-density spectrum is then $\Omega_{\rm DWD}(f)=\omega_{\rm pref}(f)[h_{\rm DWD}^{(1\,\mathrm{yr})}(f)]^2$, and conversely $S_{\rm h}(f)=\Omega_{\rm GW}(f)/\omega_{\rm pref}(f)$. This is the standard relation between a one-sided strain PSD and the dimensionless GW energy-density spectrum \cite{Phinney2001,AllenRomano1999}.

The production sky anisotropy uses the Galactic disc-plus-bulge template of Banagiri et al.~\cite{Banagiri2021}. With $R=\sqrt{x^2+y^2}$, $r=\sqrt{x^2+y^2+z^2}$, and $\widetilde r=\sqrt{x^2+y^2+(z/q_b)^2}$, the unnormalised density is
\begin{equation}
	\begin{aligned}
		\rho_{\rm DWD}(x,y,z)&=\rho_0\biggl[\exp\biggl(-\frac{R}{R_d}-\frac{|z|}{z_d}\biggr)\\
		&\quad +A_b\frac{\exp[-(r/r_{\rm cut})^2]}{[1+\widetilde r/r_b]^{\gamma}}\biggr].
	\end{aligned}
	\label{eq:dwd_spatial}
\end{equation}
The production values are $R_d=2.9\,\mathrm{kpc}$, $z_d=0.3\,\mathrm{kpc}$, $r_b=0.075\,\mathrm{kpc}$, $r_{\rm cut}=2.1\,\mathrm{kpc}$, $\gamma=1.8$, $q_b=0.5$, and $A_b=1$, with the observer at $(R_\odot,0,z_\odot)=(8.2,0,0.020)\,\mathrm{kpc}$. The response uses the line-of-sight weight $\mathcal W_{\rm DWD}(\hat k)\propto\int_0^{30\,\mathrm{kpc}}\rho_{\rm DWD}(\mathbf r_\odot+s\hat k)\,\mathrm ds$, normalized to unit quadrature-weighted mean before projection into the TianQin channels.

Rather than using the tabulated curve directly, we adopt a smooth broken-power-law parameterization as a compact phenomenological representation of the foreground spectrum. The four local-pivot coordinates make its normalization and local slope explicit at a chosen reference frequency:
\begin{equation}
	\begin{aligned}
		\Omega_{\rm DWD}(f)&=
		\Omega_{\rm ref}\,\frac{1+w_{\rm ref}}
		{1+(f/f_b)^{\Delta\alpha}}
		\left(\frac{f}{f_{\rm ref}}\right)^{\frac{s_{\rm ref}
				+\Delta\alpha\,w_{\rm ref}}{1+w_{\rm ref}}},\\
		w_{\rm ref}&=\left(\frac{f_{\rm ref}}{f_b}\right)^{\Delta\alpha}.
	\end{aligned}
	\label{eq:dwd_model}
\end{equation}
Here $f_{\rm ref}=3\,\mathrm{mHz}$, $\Omega_{\rm ref}=\Omega_{\rm DWD}(f_{\rm ref})$, and $s_{\rm ref}=\left.\mathrm d\ln\Omega_{\rm DWD}/\mathrm d\ln f\right|_{f_{\rm ref}}$. The parameter $f_b$ is the bend frequency and $\Delta\alpha$ is the change in logarithmic slope across the bend. Thus the model is normalized by construction at the reference frequency and has the sampled local slope there.

For an isotropic SGWB, the response of two TDI channels is summarized by a frequency- and time-dependent overlap-reduction function (ORF)~\cite{CornishRubbo2003,Vallisneri2005,Liang:2023fdf,Liang:2024ulf}:
\begin{equation}
	\begin{aligned}
		\Gamma_{IJ}(f,t)&=\frac{1}{8\pi}\int \dd\Omega_{\hat k}
		\left[F_I^+F_J^{+*}+F_I^\times F_J^{\times *}\right],\\
		&\qquad I,J\in\{A,E\},
	\end{aligned}
	\label{eq:orf_definition}
\end{equation}
where $F_I^+$ and $F_I^\times$ are the channel responses to the two GW polarizations, and the internal $A/E$ reference-site phases and the spacecraft propagation phase $\exp[-2\pi \mathrm{i} f\hat k\cdot\mathbf x_I(t)/c]$ are included in $F_I$ following the convention of Refs.~\cite{CornishRubbo2003,Vallisneri2005,Liang2023TDI}. Because the TianQin pointing is fixed while the constellation rotates, the projected foreground response is periodic with the orbital period $P_{\rm TQ}$. We evaluate the $AA$, $EE$, and complex $AE$ response over one period and use the resulting phase-dependent covariance in the annual likelihood.

Let $I,J\in\{A,E\}$. The total spectral covariance in phase block $k$ is
\begin{equation}
	\begin{aligned}
		C_{IJ}(f,k\mid\boldsymbol\vartheta)={}&N_{IJ}(f\mid S_{p,0},S_{a,0})\\
		&+\Gamma_{IJ}^{\rm BBH}(f,k)S_{\rm h}^{\rm BBH}(f\mid A_{\rm BBH})\\
		&+{\cal R}_{IJ}^{\rm DWD}(f,k)S_{\rm h}^{\rm DWD}(f\mid\eta_{\rm DWD}),\\[2pt]
		S_{\rm h}^{\rm DWD}(f\mid\eta_{\rm DWD})
		&=\frac{\Omega_{\rm DWD}(f\mid\eta_{\rm DWD})}{\omega_{\rm pref}(f)},\\
		\eta_{\rm DWD}&=\bigl(\log_{10}\Omega_{\rm ref},s_{\rm ref},
		\log_{10}f_b,\Delta\alpha\bigr).
	\end{aligned}
	\label{eq:cov_block}
\end{equation}
Here $k$ labels orbital-phase blocks, $f$ is Fourier frequency, and $\boldsymbol\vartheta$ denotes the sampled and fixed model parameters. The matrices $N_{IJ}$, $\Gamma_{IJ}^{\rm BBH}$, and $\mathcal R_{IJ}^{\rm DWD}$ describe instrumental noise, the isotropic BBH response, and the response-projected Galactic foreground template, respectively; $S_{\rm h}^{\rm BBH}$ and $S_{\rm h}^{\rm DWD}$ are one-sided strain PSDs. The four entries of $\eta_{\rm DWD}$ are defined in Eq.~\eqref{eq:dwd_model}. Instrumental noise is stationary with $N_{AE}=0$, and the isotropic BBH component has $\Gamma_{AE}^{\rm BBH}=0$; the complex $AE$ term of the foreground is retained. The four spectral parameters in Eq.~\eqref{eq:dwd_model} change the frequency dependence coherently across all covariance entries. We evaluate the response over one TianQin orbital period, and the phase-folding and block construction are specified in Sec.~\ref{sec:like}.

\section{Phase-folded likelihood and forecast configuration}\label{sec:like}

\subsection{Observing schedule and Fourier-mode accounting}

We specify the observing schedule, data segmentation, and frequency support used to construct the likelihood. The mock observation spans \(T_{\rm data}=365.25\,\mathrm d\) and adopts a 3+3 duty cycle consisting of two 90-day active windows separated by a 90-day gap. This schedule is an analysis assumption and is distinct from the TianQin constellation orbital period \(P_{\rm TQ}\simeq3.65\,\mathrm d\) introduced in Sec.~\ref{sec:tqdata}. The production Fourier band is \(f_{\min}=0.10175\,\mathrm{mHz}\le f\le f_{\max}=0.05\,\mathrm{Hz}\). The lower boundary is set by the first positive Fourier frequency of the elementary segment rather than by an independently chosen astrophysical cutoff; the upper boundary is the baseline analysis choice, tested separately in Sec.~\ref{sec:results}. Because the foreground spectral fit is calibrated only over \(0.5\)--\(5\,\mathrm{mHz}\), using the full band makes the inference conditional on extrapolating the fit outside that interval and on the fixed Galactic sky template. The baseline excludes an extragalactic DWD component and does not marginalize over alternative spatial templates.

The segment duration is chosen as \(T=P_{\rm TQ}/32\simeq2.73\,\mathrm h\), so that one TianQin orbital cycle is represented by 32 orbital-phase bins. This phase resolution preserves the leading periodic modulation of the Galactic foreground while retaining many independent Fourier vectors in each phase--frequency block. These 32 orbital-phase bins are distinct from the 32- and 64-point frequency-support grids compared in Sec.~\ref{sec:results}: the former discretize the time-dependent response, whereas the latter control how finely the frequency-dependent covariance is represented. The production likelihood uses 32 phase bins and 64 frequency-support points. The first positive Fourier frequency is \(1/T=0.10175\,\mathrm{mHz}\), rounded to \(0.1\,\mathrm{mHz}\) in plots.

Each 90-day active window is divided into complete segments only, so no segment crosses a 90-day off interval. With a single global orbital-phase origin at the start of the schedule, the first active window contains 791 complete segments plus a \(2052\,\mathrm s\) residual, and the second contains 790 complete segments with additional \(4104\,\mathrm s\) leading and \(6156\,\mathrm s\) trailing edges; all incomplete pieces are discarded and no zero padding is used. The complete segments are folded into 32 orbital-phase blocks, with \(n_k=49\) or \(50\) segments per block and 1581 complete segments in total, where \(n_k\) denotes the number of complete segments assigned to orbital-phase block \(k\).

For phase block \(k\) and frequency support bin \(j\), the compressed statistic is formed by summing the outer products of all independent complex \(A/E\) Fourier vectors assigned to that block and support interval. If \(m_j\) is the number of positive Fourier bins in support interval \(j\), the Wishart mode count is \(n_{kj}=n_k m_j\). The 64-point production frequency support contains the same 491 positive Fourier modes per complete segment as the 32-point validation support; changing the support grid therefore does not change the underlying Fourier-mode count. The full mode-count table is stored with the numerical results. Empty support intervals with \(m_j=0\) are omitted from the likelihood and contribute no information. The support points are response-evaluation points, not additional independent modes. Within each support interval we use the covariance evaluated at its representative frequency, so the pooled statistic is exactly Wishart only in the piecewise-constant-covariance approximation. The phase-dependent ORF and the foreground covariance are computed once per phase block and reused across cycles, while the actual \(n_{kj}\) values retain the physical Fourier-mode multiplicity.

\begin{table*}[t]
	\caption{Production population, nuisance, and physical-lensing configuration. The likelihood uses the population of PixelPop-turnover model. The downstream SIS propagation uses the same posterior draws with the full sampled mass-pair distribution compressed to 128 weighted chirp-mass nodes and the adopted SIS grid below. The quoted percentiles are forecast-posterior quantiles, not upper limits from an observed lens catalog.}
	\label{tab:physical_rates}
	\begin{ruledtabular}
		\small
		\setlength{\tabcolsep}{4pt}
		\renewcommand{\arraystretch}{1.15}
		\begin{tabular}{lll}
			\parbox[t]{0.25\textwidth}{Quantity} & \parbox[t]{0.43\textwidth}{Value or support} & \parbox[t]{0.24\textwidth}{Nodes or samples} \\ \hline
			\parbox[t]{0.25\textwidth}{GWTC-5.0 PixelPop population} & \parbox[t]{0.43\textwidth}{rate and mass distribution} & \parbox[t]{0.24\textwidth}{Ref.~\cite{LIGO2026GWTC5}} \\
			\parbox[t]{0.25\textwidth}{Population representation} & \parbox[t]{0.43\textwidth}{SGWB likelihood: \((R_0,\lambda_z,\alpha_{\rm BBH})\) plus fixed PixelPop mass moment; downstream SIS: full-mass quadrature} & \parbox[t]{0.24\textwidth}{PixelPop input; \(\alpha_{\rm BBH}\) nested} \\
			\parbox[t]{0.25\textwidth}{Population-coordinate map} & \parbox[t]{0.43\textwidth}{PixelPop rate at $z=0.2$ mapped with Eq.~\eqref{eq:rate_model}; injection 23.76; central-99\% truncated Gaussian prior in $\log_{10} R_0$, with bounds 15.40--36.65.} & \parbox[t]{0.24\textwidth}{$\alpha_z=5.6$ and $z_p=1.9$ fixed; $\lambda_z$ sampled in the turnover model.} \\
			\parbox[t]{0.25\textwidth}{Mass support} & \parbox[t]{0.43\textwidth}{entire sampled GWTC-5.0 mass range in downstream SIS propagation} & \parbox[t]{0.24\textwidth}{128 weighted chirp-mass nodes} \\
			\parbox[t]{0.25\textwidth}{BBH amplitude} & \parbox[t]{0.43\textwidth}{\(A_{\BBH}\) from Eq.~\eqref{eq:bbh_amplitude_population}} & \parbox[t]{0.24\textwidth}{derived, not sampled} \\
			\parbox[t]{0.25\textwidth}{Continuous foreground coordinates} & \parbox[t]{0.43\textwidth}{\((\log_{10}\Omega_{\rm ref},s_{\rm ref},\log_{10}f_b,\Delta\alpha)\)} & \parbox[t]{0.24\textwidth}{nested sampling} \\
			\parbox[t]{0.25\textwidth}{Noise PSD coordinates} & \parbox[t]{0.43\textwidth}{\((\log_{10}S_{p,0},\log_{10}S_{a,0})\)} & \parbox[t]{0.24\textwidth}{nested sampling} \\
			\parbox[t]{0.25\textwidth}{Wishart sampler} & \parbox[t]{0.43\textwidth}{\(n_{\rm live}=1024\), \(\Delta\log Z=0.05\)} & \parbox[t]{0.24\textwidth}{50 realizations} \\
			\parbox[t]{0.25\textwidth}{SIS \((z_s,z_l,\sigma_v,y,\mathrm{orientation})\)} & \parbox[t]{0.43\textwidth}{cached physical quadrature} & \parbox[t]{0.24\textwidth}{\((64,48,24,48,128)\)} \\
			\parbox[t]{0.25\textwidth}{Frequency support} & \parbox[t]{0.43\textwidth}{\(0.10175\,\mathrm{mHz}\)--\(0.05\,\mathrm{Hz}\)} & \parbox[t]{0.24\textwidth}{64 support points} \\
		\end{tabular}
	\end{ruledtabular}
\end{table*}

\subsection{Complex Wishart likelihood}

For a phase block \(k\) and frequency support bin \(j\), let \(\mathbf d_{kjr}\) denote the two-component complex Fourier vectors, with \(r=1,\dots,n_{kj}\) indexing the independent vectors in that block. The normalized sample spectral matrix is the empirical covariance
\begin{equation}
	\widehat{\mathbf C}_{kj}=\frac{1}{n_{kj}}\sum_{r=1}^{n_{kj}}\mathbf d_{kjr}\mathbf d_{kjr}^{\dagger},
	\label{eq:sample_cov}
\end{equation}
where \(\dagger\) denotes the complex-conjugate transpose. If the Fourier coefficients are zero-mean complex Gaussian, \(\widehat{\mathbf C}_{kj}\) follows a \(2\times2\) complex Wishart distribution \cite{Wishart1928,Goodman1963}. Up to parameter-independent constants, the corresponding log-likelihood contribution is
\begin{equation}
	\ln\mathcal L_{W,kj}=-n_{kj}\left[\ln\det\mathbf\Sigma_{kj}(\boldsymbol\vartheta)+\operatorname{tr}\left(\mathbf\Sigma_{kj}^{-1}(\boldsymbol\vartheta)\widehat{\mathbf C}_{kj}\right)\right],
	\label{eq:wishart}
\end{equation}
with \(\mathbf\Sigma_{kj}\) given by Eq.~\eqref{eq:cov_block}.

The full likelihood is \(\mathcal L_W=\prod_{k,j}\mathcal L_{W,kj}\), and the covariance is recomputed at every population and foreground parameter point. The smallest nonempty folded block has \(n_{kj}\geq49\), so the regular \(p=2\) Wishart density is applicable to every retained support interval; empty support intervals are omitted as described above. Cholesky factorization is used for the determinant and inverse, and non-positive-definite trial covariances are rejected.

The folded likelihood is equivalent, up to parameter-independent terms, to the product of the corresponding complex-Gaussian Fourier kernels. It retains the diagonal \(AA\) and \(EE\) powers and the complex \(AE\) correlation, while avoiding repeated evaluation of nearly identical response matrices over the 3.64-day cycle.

\subsection{Production configuration and priors}

The nine production coordinates are summarized in Table~\ref{tab:physical_rates}. They comprise the three BBH population coordinates \((R_0,\lambda_z,\alpha_{\rm BBH})\), the four foreground coordinates \((\log_{10}\Omega_{\rm ref},s_{\rm ref},\log_{10}f_b,\Delta\alpha)\), and the two instrumental-noise coordinates \((\log_{10}S_{p,0},\log_{10}S_{a,0})\). The SIS lens, image-selection, and association ingredients are fixed, and \(A_{\BBH}\) is derived from the population integral in Eq.~\eqref{eq:bbh_amplitude_population} rather than sampled independently; the sampled BBH spectral coordinate is the tilt \(\alpha_{\rm BBH}\).

The population input is the merger rate and mass distribution from the PixelPop model of GWTC-5.0, the production representation evaluates Eq.~\eqref{eq:rate_model} with fixed \(\alpha_z=5.6\) and \(z_p=1.9\), while \(\lambda_z\) remains sampled. We adopt a flat-\(\Lambda\)CDM cosmology with \((H_0,\Omega_m)=(67.4\,\mathrm{km\,s^{-1}\,Mpc^{-1}},0.315)\) and integrate the source population to \(z_{\max}=10\). The physical propagation uses the SIS velocity-dispersion function in Eq.~\eqref{eq:sis_vdf}, the two-image region \(0<y<1\), the detector-frame chirp-mass scaling in Eq.~\eqref{eq:image_snr}, the Marcum-\(Q\) selection probability, and unit association efficiency. The canonical BBH amplitude pivot is \(f_0=3\,\mathrm{mHz}\). The production likelihood covers \(0.10175\,\mathrm{mHz}\)--\(0.05\,\mathrm{Hz}\) with 64 logarithmic response-support points and physical Fourier-bin weights.

The six foreground/noise coordinates have Gaussian prior widths \((0.40,0.35,0.40,0.50,0.25,0.25)\), in the order listed in Table~\ref{tab:physical_rates}, while \(\alpha_{\rm BBH}\) has prior mean \(2/3\) and standard deviation \(0.6\). The likelihood samples the foreground, SGWB, and instrumental-noise coordinates continuously with nested sampling. The reported system fractions are posterior-predictive quantiles conditional on the fixed SIS and search model.

\begin{table*}[t]
	\caption{Conditional single-TianQin SIS forecast. The entries are quantiles of the equal-weight aggregate mixture of 50 full-mass rate propagations on the adopted SIS grid using the posterior of PixelPop-turnover model. They are forecast posterior percentiles, not upper limits inferred from an observed lens catalogue. The event fraction uses the all-event denominator in Eq.~\eqref{eq:event_fraction}; the system fraction uses the one-source/one-system denominator in Eq.~\eqref{eq:system_fraction}. The headline rate values in the text use the corresponding pooled-mixture medians.}
	\label{tab:physical_forecast}
	\begin{ruledtabular}
		\small
		\begin{tabular}{lccc}
			Quantity & $\mu_0=2$ & $\mu_0=5$ & $\mu_0=10$ \\ \hline
			\(R_{\rm event}^{\rm det}\), posterior median & \(5.50\times10^{-4}\) & \(4.93\times10^{-4}\) & \(4.56\times10^{-4}\) \\
			\(R_{\rm sys}^{\rm det}\), posterior median & \(2.54\times10^{-4}\) & \(2.53\times10^{-4}\) & \(2.47\times10^{-4}\) \\
			\(\mathcal F_{\rm event}^{\rm det}\), posterior median & \(7.24\times10^{-4}\) & \(6.49\times10^{-4}\) & \(6.01\times10^{-4}\) \\
			\(\mathcal F_{\rm event}^{\rm det}\), posterior 90th percentile & \(1.01\times10^{-3}\) & \(9.23\times10^{-4}\) & \(8.65\times10^{-4}\) \\
			\(\mathcal F_{\rm sys}^{\rm det}\), posterior median & \(3.35\times10^{-4}\) & \(3.33\times10^{-4}\) & \(3.25\times10^{-4}\) \\
			\(\mathcal F_{\rm sys}^{\rm det}\), posterior 90th percentile & \(4.78\times10^{-4}\) & \(4.75\times10^{-4}\) & \(4.66\times10^{-4}\) \\
		\end{tabular}
	\end{ruledtabular}
\end{table*}

\subsection{Mock data and implementation checks}\label{sec:sim}

We generate 50 independent mock realizations comprising the injected SGWB from BBH, the response-projected Galactic foreground template, and stationary instrumental noise. Each realization is analyzed with the nine-coordinate continuous local-pivot model described above, and the resulting posterior samples are propagated through the cached SIS image-selection kernel. We numerically verify the positive definiteness of the covariance, the phase-folded Wishart kernel, the zero-optical-depth and association-efficiency limits, and the Marcum-\(Q\) interpolation. The convergence and finite-ensemble coverage results are reported in Sec.~\ref{sec:results}.

\begin{figure}[t]
	\IfFileExists{tianqin_noise_parameter_posterior_ensemble.png}{%
		\includegraphics[width=0.95\columnwidth]{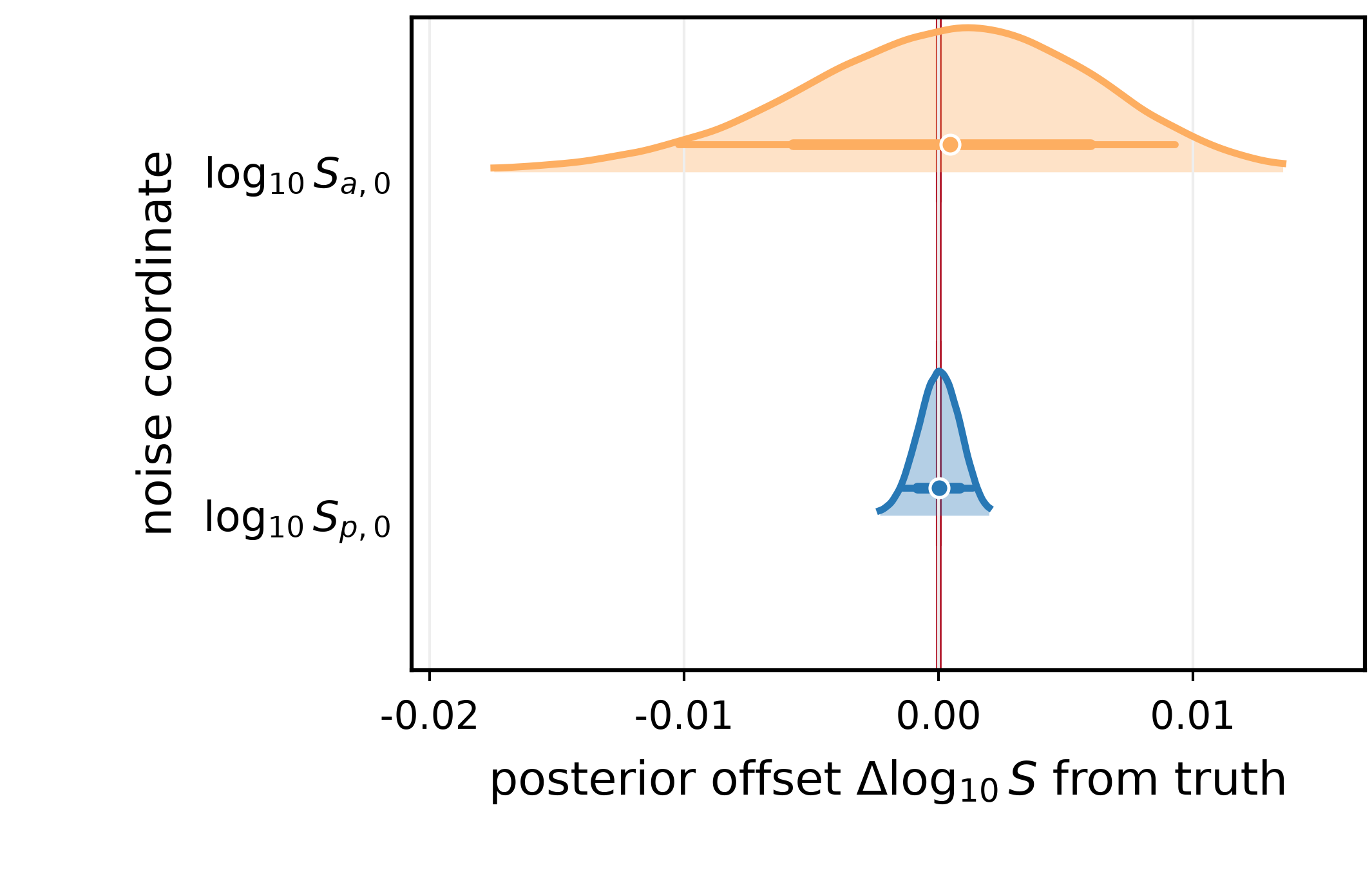}%
	}{\fbox{\parbox{0.95\columnwidth}{The noise-parameter posterior figure is included in the full project build.}}}
	\caption{Horizontal posterior-ridge display for the two instrumental-noise PSD coordinates from the 50 50-realization Wishart ensemble. The filled ridges show the normalized one-dimensional mixtures, thin and thick bars show the 5--95\% and 16--84\% intervals, points mark posterior medians, and the red vertical line marks zero offset from truth.}
	\label{fig:noise_posterior}
\end{figure}

\begin{figure*}[t]
	\IfFileExists{reference_style_posterior.png}{
		\includegraphics[width=0.99\textwidth]{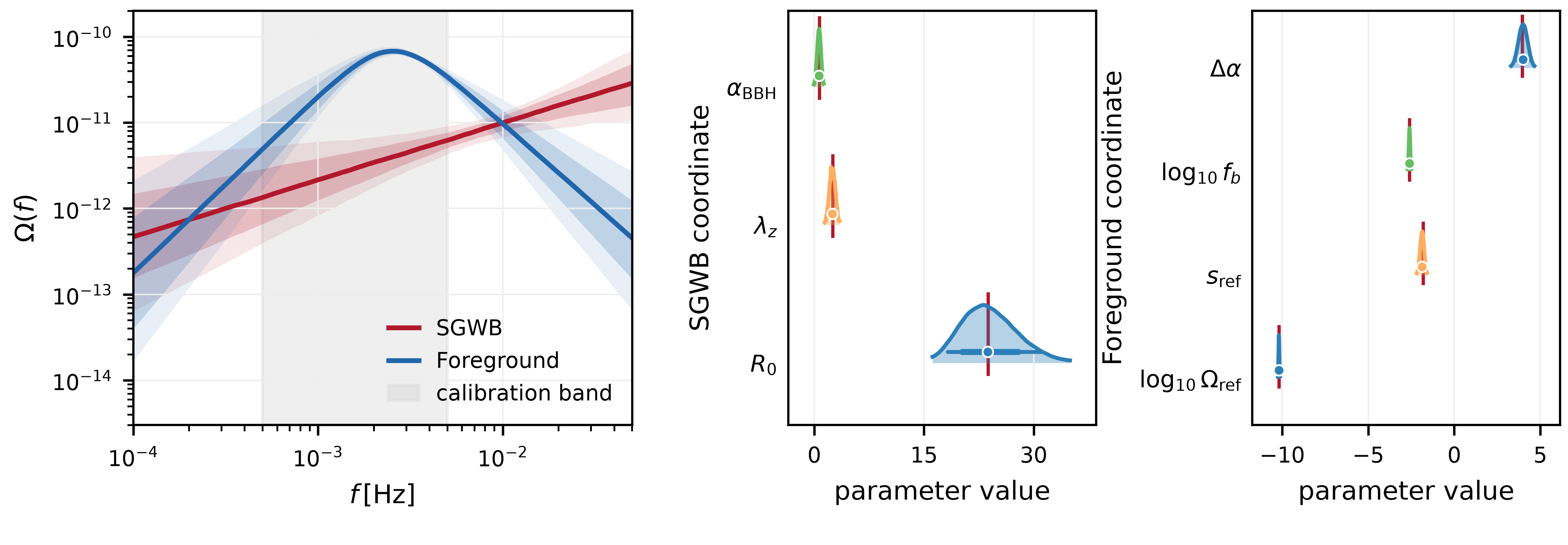}%
	}{\fbox{\parbox{0.95\textwidth}{The foreground diagnostic figure is included in the full project build.}}}
	\caption{Three-panel posterior diagnostic for the production ensemble. (a) Equal-weight posterior bands for the SGWB and Galactic foreground. Solid curves show posterior medians, and the dark and light bands enclose the central 68\% and 90\% credible intervals, respectively; the gray strip marks the adopted calibration band. (b) One-dimensional marginal posteriors of the three parameters $(R_0,\lambda_z,\alpha_{\rm BBH})$ for the SGWB. (c) One-dimensional marginal posteriors of the four parameters $(\log_{10}\Omega_{\rm ref},s_{\rm ref},\log_{10}f_b,\Delta\alpha)$ for the foreground. In (b) and (c), red vertical lines mark the injected truth, and the same 68\% and 90\% credible-interval conventions are used.}
	\label{fig:foreground_diagnostic}
\end{figure*}

\begin{figure}[t]
	\includegraphics[width=0.98\columnwidth]{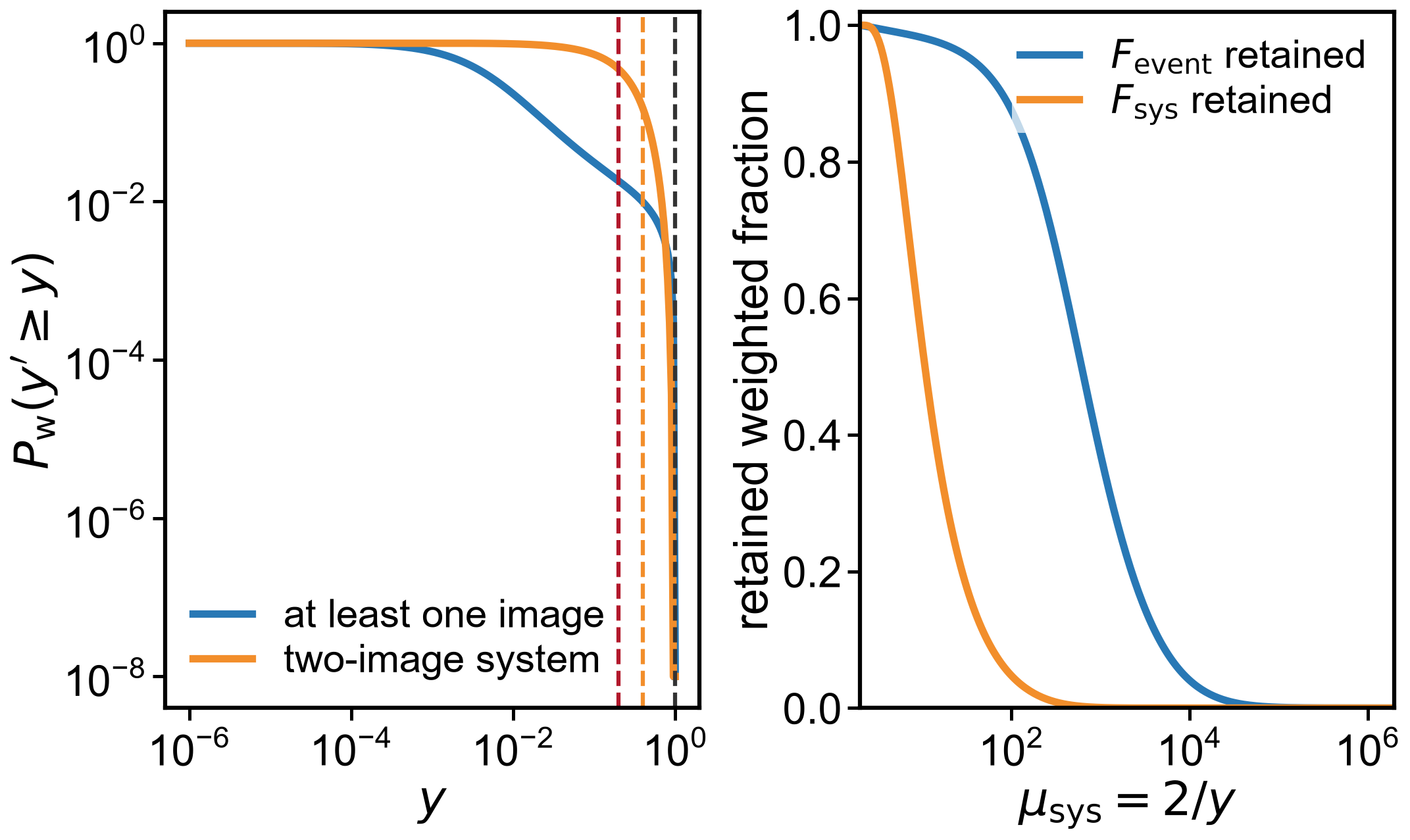}
	\caption{\label{fig:sis_mu_distribution}Diagnostic for the SIS magnification-threshold implementation. The curves show the normalized event- and system-weighted distributions of $\mu_{\rm sys}$, together with the cumulative fractions retained above $\mu_0=2,5,10$. The threshold relation $\mu_{\rm sys}=2/y$ is imposed before the event and system kernels are accumulated.}
\end{figure}

\section{Rate forecasts and validation}\label{sec:results}

\subsection{Posterior-predictive lens rates and event counts}
The posterior-predictive rate and fraction summaries are listed in Table~\ref{tab:physical_forecast}. the nine-coordinate posterior mixture and spectral diagnostics are shown in Figs.~\ref{fig:noise_posterior} and~\ref{fig:foreground_diagnostic}, and Fig.~\ref{fig:sis_mu_distribution} checks the SIS magnification-threshold implementation. 
With $\alpha_{\rm BBH}$ sampled, the SGWB constrains both the population-induced pivot amplitude and a spectral tilt, while the posterior-predictive rates retain dependence on the redshift evolution and mass model. We therefore propagate the full population posterior through the same SIS and image-selection calculation for every sample, preserving the correlation between all predicted rates.

For each quantity, we define the 16th--84th percentile width reduction as
\begin{equation}
	\Delta_w=1-\frac{q_{84}^{\rm post}-q_{16}^{\rm post}}
	{q_{84}^{\rm prior}-q_{16}^{\rm prior}},
\label{eq:width_reduction}
\end{equation}
using the unrounded interval endpoints from the numerical output; the endpoints quoted in the text and tables are rounded for readability. These reductions are conditional on the adopted population, SIS lens, search, association, covariance, and foreground models; they are not model-independent measurement precisions.

For clarity, the quoted information gains are computed from the interval widths, not from a change in the medians. At $\mu_{\rm sys}\ge2$, the prior-only and pooled-posterior intervals (in units of $10^{-4}$) are $(3.88,7.92)$ and $(3.94,7.71)$ for $R_{\rm event}^{\rm det}$, $(1.76,3.73)$ and $(1.79,3.63)$ for $R_{\rm sys}^{\rm det}$, and $(2.56,4.51)$ and $(2.55,4.43)$ for $\mathcal F_{\rm sys}^{\rm det}$. Substitution into Eq.~(\ref{eq:width_reduction}) gives $1-(7.71-3.94)/(7.92-3.88)=6.50\%$, $1-(3.63-1.79)/(3.73-1.76)=6.42\%$, and $1-(4.43-2.55)/(4.51-2.56)=3.95\%$, which are reported as 6.5\%, 6.4\%, and 4.0\%, respectively.

The event rate exceeds the system rate because the latter requires both images to exceed the threshold and to lie within the association time window. For an SIS lens, $\mu_{\rm sys}=2/y$ in the two-image region, so the threshold mask is exactly $y\le 2/\mu_0$. We verified this identity directly in the cached kernel; the corresponding event/system weighted retained fractions are $0.99/0.85$ at $\mu_0=5$ and $0.98/0.52$ at $\mu_0=10$. The two-image selection therefore concentrates the system numerator toward small $y$, while the lensed contribution remains a small correction to the all-detected denominator. The near equality of the three system-fraction columns in Table~\ref{tab:physical_forecast} is consequently a property of the ratio and selection weighting, rather than a threshold-cache error.

\subsection{Numerical convergence, coverage, and model-systematic checks}

\emph{Unit checks and phase-folded likelihood.} The system-rate denominator passes three unit checks. Setting the SIS optical depth to zero gives \(R_{\rm all}^{\rm det}=R_{\rm unlensed}^{\rm det}\), the baseline calculation satisfies \(R_{\rm all}^{\rm det}=R_{\rm unlensed}^{\rm det}+R_{\rm lensed,sys}^{\rm det}\) to numerical precision, and setting the association efficiency to zero eliminates \(R_{\rm sys}^{\rm det}\) while leaving the at-least-one-image event rate nonzero. These checks verify the one-source/one-system counting rule used in Eq.~\eqref{eq:system_fraction}. The phase-folded TianQin likelihood uses \(T=P_{\rm TQ}/32\) and retains the actual Fourier mode count in every phase-frequency block. The two 90-day windows contribute 791 and 790 complete segments; incomplete pieces are discarded. The smallest nonempty \(2\times2\) Wishart block contains at least 49 independent vectors, so the regular complex Wishart density is applicable throughout the production likelihood.

\emph{Numerical convergence.} The final ensemble adopts the corrected SIS velocity-dispersion normalization and the cached rate kernel with $(n_z,n_{z_l},n_{\sigma},n_y,n_{\rm orient})=(64,48,24,48,128)$, using the same image-level Marcum-\(Q\) selection for every posterior sample. To assess numerical convergence, we repeated the SIS propagation for the same 50 posterior realizations on three grids, $(64,48,24,48,128)$, $(96,72,36,72,256)$, and $(128,96,48,96,384)$, referred to below as grids A, B, and C, respectively, all with a common nested Sobol orientation sequence. The B-to-C changes in the ensemble-median posterior medians and posterior-\(q_{90}\) values are below 1.7\% for both rates and below 0.6\% for \(\mathcal F_{\rm sys}^{\rm det}\) at all three thresholds. Relative to grid A, grid C shifts the rate summaries by 2.1--2.2\% and the system-fraction summaries by 0.6--0.7\%. Balancing numerical accuracy against computational cost, we adopt grid A as the final SIS configuration.

\emph{Frequency pooling.} We repeated the matched 50-seed local-pivot diagnostic with 32 and 64 logarithmic support points, retaining the same 32 orbital-phase blocks, PixelPop-turnover population, injection, and nominal PSD. The ensemble median changes from 32 to 64 points are $0.20$--$0.29\%$ for $R_{\rm event}^{\rm det}$, $0.27$--$0.30\%$ for $R_{\rm sys}^{\rm det}$, and $0.49\%$ for $\mathcal F_{\rm sys}^{\rm det}$. The 64-point support remains the canonical production configuration.

\begin{table*}[t]
	\caption{Discrete-nuisance frequency-pooling validation for the current PixelPop-turnover population. Each entry gives the 16th, 50th, and 84th percentiles across the 50 realization-specific posterior \(q_{90}\) values.}
	\label{tab:frequency_ensemble}
	\renewcommand{\arraystretch}{1.08}\small
	\begin{ruledtabular}
		\begin{tabular}{lc ccc ccc c}
			Quantity & $\mu_0$ & \multicolumn{3}{c}{32 support points} & \multicolumn{3}{c}{64 support points} & $\Delta_{64/32}$\\
			& & $q_{16}$ & $q_{50}$ & $q_{84}$ & $q_{16}$ & $q_{50}$ & $q_{84}$ & \\ \hline
			$R_{\rm event}^{\rm det}$ & 2 & $9.21\times10^{-5}$ & $1.03\times10^{-4}$ & $1.14\times10^{-4}$ & $9.76\times10^{-5}$ & $1.03\times10^{-4}$ & $1.11\times10^{-4}$ & $+0.20\%$\\
			$R_{\rm sys}^{\rm det}$ & 2 & $4.77\times10^{-5}$ & $5.37\times10^{-5}$ & $5.95\times10^{-5}$ & $5.07\times10^{-5}$ & $5.38\times10^{-5}$ & $5.80\times10^{-5}$ & $+0.27\%$\\
			$\mathcal F_{\rm sys}^{\rm det}$ & 2 & $2.41\times10^{-4}$ & $2.74\times10^{-4}$ & $3.02\times10^{-4}$ & $2.58\times10^{-4}$ & $2.75\times10^{-4}$ & $2.96\times10^{-4}$ & $+0.49\%$\\
			$R_{\rm event}^{\rm det}$ & 5 & $8.97\times10^{-5}$ & $1.01\times10^{-4}$ & $1.12\times10^{-4}$ & $9.52\times10^{-5}$ & $1.01\times10^{-4}$ & $1.09\times10^{-4}$ & $+0.21\%$\\
			$R_{\rm sys}^{\rm det}$ & 5 & $4.77\times10^{-5}$ & $5.36\times10^{-5}$ & $5.95\times10^{-5}$ & $5.06\times10^{-5}$ & $5.38\times10^{-5}$ & $5.79\times10^{-5}$ & $+0.30\%$\\
			$\mathcal F_{\rm sys}^{\rm det}$ & 5 & $2.41\times10^{-4}$ & $2.74\times10^{-4}$ & $3.02\times10^{-4}$ & $2.57\times10^{-4}$ & $2.75\times10^{-4}$ & $2.95\times10^{-4}$ & $+0.49\%$\\
			$R_{\rm event}^{\rm det}$ & 10 & $8.82\times10^{-5}$ & $9.92\times10^{-5}$ & $1.10\times10^{-4}$ & $9.37\times10^{-5}$ & $9.95\times10^{-5}$ & $1.07\times10^{-4}$ & $+0.29\%$\\
			$R_{\rm sys}^{\rm det}$ & 10 & $4.74\times10^{-5}$ & $5.34\times10^{-5}$ & $5.92\times10^{-5}$ & $5.04\times10^{-5}$ & $5.35\times10^{-5}$ & $5.76\times10^{-5}$ & $+0.29\%$\\
			$\mathcal F_{\rm sys}^{\rm det}$ & 10 & $2.39\times10^{-4}$ & $2.72\times10^{-4}$ & $3.01\times10^{-4}$ & $2.56\times10^{-4}$ & $2.74\times10^{-4}$ & $2.94\times10^{-4}$ & $+0.49\%$\\
		\end{tabular}
	\end{ruledtabular}
\end{table*}

\emph{Posterior coverage.} Using the independent SIS truth generated with the same PixelPop-turnover model, the realization-specific 68\% intervals contain the truth in 48/50 cases for both $R_{\rm event}^{\rm det}$ and $R_{\rm sys}^{\rm det}$ at each \(\mu_{\rm sys}=2,5,10\); the corresponding 90\% intervals contain the truth in 50/50 cases. For $\mathcal F_{\rm sys}^{\rm det}$, the counts are 50/50 for both nominal levels at all three thresholds. These finite-ensemble counts are validation diagnostics rather than a claim of exact frequentist coverage.

\emph{Response and foreground-model sensitivity.} We performed three compact sensitivity checks for the response-projected DWD foreground. Increasing the production DWD sky quadrature from \((n_\mu,n_\varphi)=(64,128)\) to \((96,192)\) changes the \(AA\) and \(EE\) covariance components by at most \(0.28\%\) and the complex \(AE\) component by at most \(1.30\%\) over eight orbital phases and eight frequencies. Independently, the \(4\times8\) isotropic ORF used by the single-TianQin auto likelihood differs from a \(32\times64\) reference by a Frobenius relative error of \(1.7\times10^{-9}\) and a maximum scaled pointwise difference of \(1.7\times10^{-5}\). A controlled analytic-response perturbation, consisting of a \(1\%\) diagonal amplitude change and a \(0.02\)-rad \(AE\) phase shift, changes the projected covariance by at most \(2.0\%\). These tests support the adopted foreground and isotropic auto-ORF quadratures at the level of this forecast. The compact local-pivot foreground spectrum is calibrated to the one-year TianQin foreground fit over \(0.5\)--\(5\,\mathrm{mHz}\). Within that interval, its relative deviation from the polynomial reference has median \(4.8\%\), 95th percentile \(8.6\%\), and maximum \(25.8\%\) on the audit grid. Extending the comparison to \(0.3\)--\(10\,\mathrm{mHz}\) produces substantially larger discrepancies because the fitted form is extrapolated outside its calibration interval. The resulting system-rate forecasts are therefore conditional on this foreground extrapolation; the calculation does not claim a calibrated foreground model over the full production band.

The production analysis keeps the full phase-dependent complex covariance and the $A/E$ cross spectrum. The association rule counts each source once and requires both SIS images to satisfy the adopted search and association conditions. The updated forecast keeps the association efficiency fixed at unity and uses the same SIS velocity-dispersion function in the prior-only and posterior propagations; the reported uncertainty is therefore conditional on this lens and search model. 

\emph{Association and SIS-normalization sensitivity.} The event fraction is unchanged by the association window. At $\mu_{\rm sys}\ge2$, the rate-weighted removed fraction is $3.81\times10^{-4}$ for a 30-day gate and negligible for 180 and 365.25 days. Setting $\epsilon_{\rm assoc}=0.5$ halves the system branch while leaving the event branch unchanged. Scaling $\phi_\ast$ by 0.8 or 1.2 rescales the lensing rates and fractions.

\begin{table}[t]
	\caption{Sensitivity of the \(\mu_{\rm sys}\ge2\) posterior-predictive fractions to the association and SIS-population inputs for the current turnover population. Entries are quoted as \(10^{-3}\times(\mathrm{median}/q_{90})\).}
	\label{tab:association_sensitivity}
	\renewcommand{\arraystretch}{1.08}\small
	\begin{ruledtabular}
		\begin{tabular}{@{}lcc@{}}
			Case & Event fraction & System fraction\\ \hline
			\(T_{\rm assoc}=30\,\mathrm d\) & \(0.369/0.538\) & \(0.191/0.281\)\\
			\(T_{\rm assoc}=180\,\mathrm d\) & \(0.369/0.538\) & \(0.191/0.281\)\\
			\(T_{\rm assoc}=365.25\,\mathrm d\) & \(0.369/0.538\) & \(0.191/0.281\)\\
			\(\epsilon_{\rm assoc}=0.5\) & \(0.369/0.538\) & \(0.095/0.140\)\\
			\(\phi_\ast=0.8\,\phi_\ast^{\rm ref}\) & \(0.295/0.430\) & \(0.153/0.225\)\\
			\(\phi_\ast=1.2\,\phi_\ast^{\rm ref}\) & \(0.442/0.645\) & \(0.229/0.337\)\\
		\end{tabular}
	\end{ruledtabular}
\end{table}

\emph{Ablation of phase and cross-spectrum information.} For the current injection, the full, phase-averaged, and diagonal likelihoods give $\mathcal F_{\rm sys}^{\rm det}(2)$ median/90th-percentile values of $1.72\times10^{-4}/2.51\times10^{-4}$, $1.75\times10^{-4}/2.52\times10^{-4}$, and $1.74\times10^{-4}/2.48\times10^{-4}$, respectively. The same 32 phase blocks, $0.05\,\mathrm{Hz}$ band, Wishart realization, and $n_{\rm live}=256$ are used in all three cases; this is a one-realization diagnostic.

\emph{Upper frequency boundary.} Extending the upper boundary from $0.05$ to $0.1\,\mathrm{Hz}$ changes $\mathcal F_{\rm sys}^{\rm det}(2)$ from $1.72\times10^{-4}$ to $1.71\times10^{-4}$ in the median ($-0.63\%$), and from $2.51\times10^{-4}$ to $2.46\times10^{-4}$ at $q_{90}$ ($-1.83\%$). The corresponding $q_{90}$ changes are $-2.17\%$ for $R_{\rm event}^{\rm det}$ and $-2.38\%$ for $R_{\rm sys}^{\rm det}$. The lower boundary is $0.10175\,\mathrm{mHz}$, with 776271 and 1552542 positive modes in the two bands, $0.05\,\mathrm{Hz}$ therefore remains the canonical choice.

The small downward shifts in Table~\ref{tab:frequency_sensitivity} do not represent a physical decrease of the merger rate with frequency. Rather, they reflect a change in likelihood truncation: extending $f_{\max}$ adds high-frequency Fourier modes whose SGWB and foreground contributions are relatively weak compared with the instrumental covariance, but which still reweight the correlated population and nuisance posterior. The resulting change in the posterior-predictive lensing rates is therefore a frequency-support and covariance-weighting effect, and its magnitude (below 2.4\% for the quoted 90th-percentile changes) is the relevant robustness statement.

\begin{table}[t]
	\caption{Paired frequency-band sensitivity for the current representative realization. Entries are posterior medians (\(q_{50}\)) and posterior 90th percentiles (\(q_{90}\)) from the SIS propagation. The canonical production band is \(0.10175\,\mathrm{mHz}\)--\(0.05\,\mathrm{Hz}\), and the sensitivity run extends only the upper boundary to \(0.1\,\mathrm{Hz}\).}
	\label{tab:frequency_sensitivity}
	\renewcommand{\arraystretch}{1.12}\small
	\begin{ruledtabular}
		\begin{tabular}{lccc}
			Quantity & $\mu_0$ & $0.05\,\mathrm{Hz}$ & $0.1\,\mathrm{Hz}$\\ \hline
			\(\mathcal F_{\rm sys}^{\rm det}\) (\(q_{50}\)) & 2 & \(1.72\times10^{-4}\) & \(1.71\times10^{-4}\)\\
			\(\mathcal F_{\rm sys}^{\rm det}\) (\(q_{90}\)) & 2 & \(2.51\times10^{-4}\) & \(2.46\times10^{-4}\)\\
			\(R_{\rm event}^{\rm det}\) (\(q_{90}\)) & 2 & \(9.77\times10^{-5}\) & \(9.56\times10^{-5}\)\\
			\(R_{\rm sys}^{\rm det}\) (\(q_{90}\)) & 2 & \(5.09\times10^{-5}\) & \(4.97\times10^{-5}\)\\
		\end{tabular}
	\end{ruledtabular}
\end{table}

\section{Summary}\label{sec:summary}

We have constructed a single-TianQin posterior-predictive framework that connects a measurement of the millihertz SGWB from BBH mergers to image-selected event rates and associated lens-system rates. The population connection extends the source-level SGWB--lensing framework developed for ground-based observations \cite{Buscicchio2020,Mukherjee2020Lensing} by retaining the phase-dependent $A/E$ covariance, modeling the Galactic DWD foreground within the same covariance, and applying image-level selection while counting lensed source events and associated lens systems as distinct observables. The resulting observable is therefore a system rate, rather than an image count or an optical depth multiplied by a source rate.

For the PixelPop-turnover configuration, the pooled 50-run full-mass SIS propagation yields \(R_{\rm event}^{\rm det}(\mu_{\rm sys}\ge2)=5.50\times10^{-4}\,\mathrm{yr}^{-1}\) and \(R_{\rm sys}^{\rm det}(\mu_{\rm sys}\ge2)=2.54\times10^{-4}\,\mathrm{yr}^{-1}\). The pooled-mixture medians are \(\mathcal F_{\rm event}^{\rm det}(2)=7.24\times10^{-4}\) and \(\mathcal F_{\rm sys}^{\rm det}(2)=3.35\times10^{-4}\), with 90th percentiles of \(1.01\times10^{-3}\) and \(4.78\times10^{-4}\), respectively. Relative to the prior-only turnover propagation, the pooled 16--84\% width reductions are \(6.5\%\), \(6.4\%\), and \(4.0\%\) for the event rate, system rate, and system fraction, respectively. These reductions are conditional on the GWTC-5.0 PixelPop mass projection, the fixed turnover parameters \(\alpha_z=5.6\) and \(z_p=1.9\), the DWD spatial template, the SIS lens model, the image search and association rules, and the piecewise-constant covariance approximation; they do not represent model-independent measurement precision.

The inference has a clear identifiability boundary. With $\alpha_{\rm BBH}$ sampled, the SGWB constrains both the population-induced pivot amplitude and a spectral tilt, but the strong correlation between $\lambda_z$ and $\alpha_{\rm BBH}$ leaves the redshift history only partially identified. The rates above are therefore posterior-predictive forecasts under the GWTC-5.0 population, a fixed SIS velocity-dispersion function, the optimal-SNR calibration, the Marcum-\(Q\) image-selection model, a one-year association window, and $\epsilon_{\rm assoc}=1$. The distinction between source-level fractions and system-level rates is also essential when comparing with earlier lensing-background analyses \cite{Buscicchio2020,Oguri2018,Mukherjee2020Lensing}.

The validation is finite-ensemble and model-conditional. The response and foreground audits support the adopted DWD sky quadrature, isotropic auto-ORF, and 0.5--5 mHz foreground calibration band, while extrapolation outside that band remains a systematic limitation. The calculation does not marginalize over the SIS population, the ground-based search efficiency, the association efficiency, the delay window, the DWD sky model, or the analytic orbital response.

The main result is a conditional posterior-propagation framework in which a phase-resolved likelihood can be connected to physically defined detectable-lens events while preserving the distinction between image, event, and system counts. It provides a quantitative baseline for future analyses that replace the fixed lens and search ingredients with empirically calibrated population and association models.

\begin{acknowledgments}
This work has been supported by the National Key Research and Development Program of China (No. 2023YFC2206704). Z.C.L. is supported by the Natural Science Foundation of Henan (Grant No. 262300422640).
\end{acknowledgments}

\appendix
\section{Likelihood checks and joint posterior}\label{app:wishart}

\begin{figure*}[t]
	\IfFileExists{cornerplot.png}{%
		\includegraphics[width=0.98\textwidth]{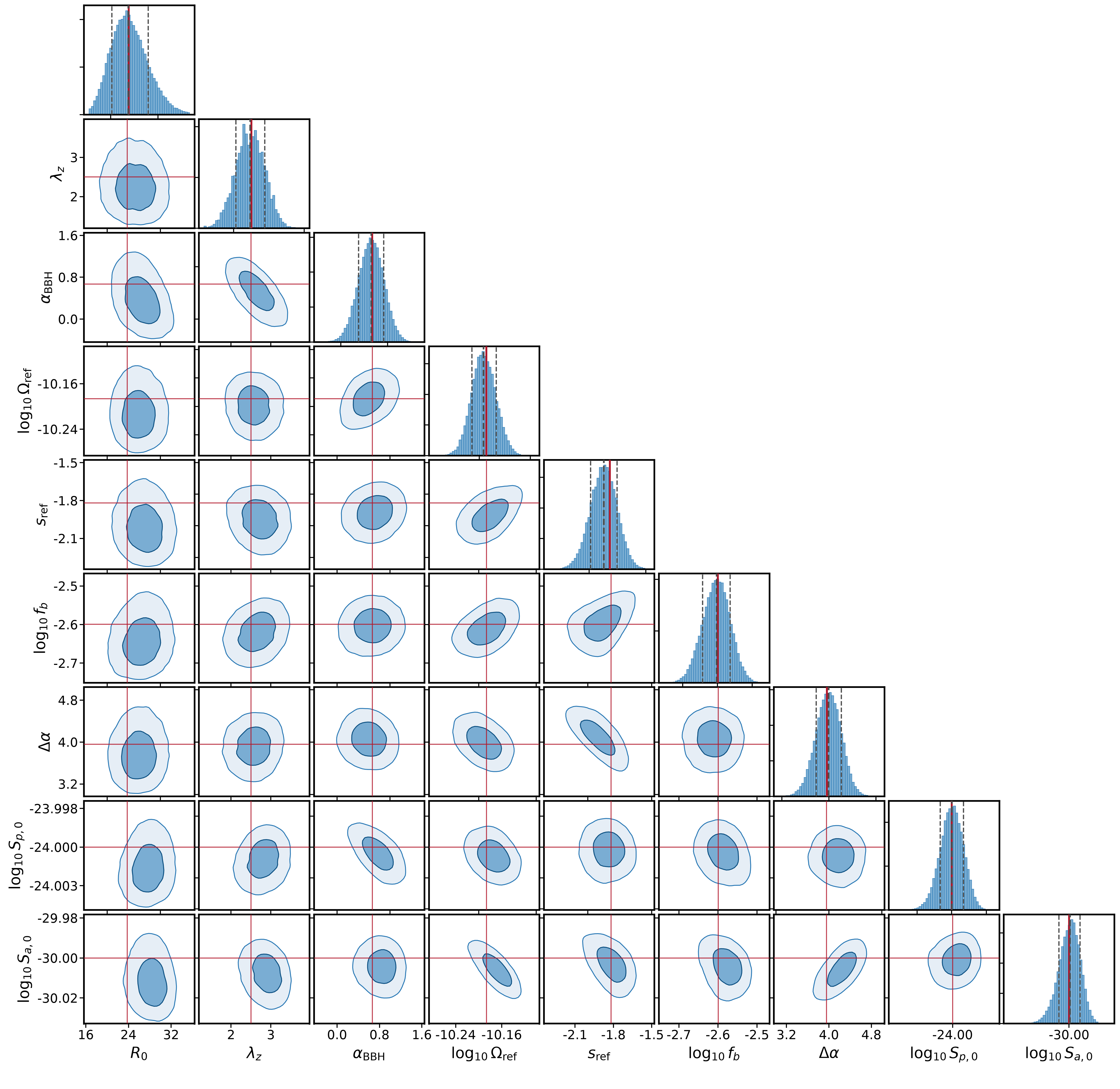}%
	}{\fbox{\parbox{0.95\textwidth}{The full corner plot is included in the full project build.}}}
	\caption{\label{app:full_corner}Nine-coordinate equal-weight posterior mixture from 50 realizations. Red solid lines mark the injected truth, while vertical dashed lines on each one-dimensional posterior mark the 16 percent, 50 percent, and 84 percent quantiles.}
\end{figure*}

The resulting nine-dimensional joint posterior is summarized in Fig.~\ref{app:full_corner}. To validate the likelihood implementation, we first consider a $p$-component complex Gaussian Fourier vector with covariance $\mathbf\Sigma$. Its sample covariance, defined in Eq.~\eqref{eq:sample_cov}, follows the complex Wishart density
\begin{equation}
	p(\widehat{\mathbf C}|\mathbf\Sigma)\propto \frac{|\widehat{\mathbf C}|^{n-p}}{|\mathbf\Sigma|^n}\exp\left[-n\,\operatorname{tr}(\mathbf\Sigma^{-1}\widehat{\mathbf C})\right],
	\label{eq:wishart_density}
\end{equation}
for $n\ge p$. We generate complex Gaussian Fourier vectors with the model covariance, form sample matrices, and compare their empirical covariance with the analytic prediction of Eq.~\eqref{eq:wishart_density}. The determinant and all principal minors are checked numerically, and trial points that fail the positive-definiteness test are assigned zero likelihood.

The production calculation uses a $2\times2$ Hermitian spectral matrix for the TianQin $A/E$ channels. For the adopted piecewise-constant support covariance, the phase-folded kernel agrees with the product of the underlying complex-Gaussian Fourier kernels up to parameter-independent terms. Tests around $T=P_{\rm TQ}/32$ confirm the trade-off between phase resolution and the number of independent Fourier vectors, and the exact phase-folded choice is adopted for the main result. 

\footnotesize
\bibliographystyle{apsrev4-1}
\bibliography{ref}

\end{document}